\documentstyle[aps,manuscript,epsf]{revtex}
\def\bm{\begin{mathletters}}
\def\em{\end{mathletters}}
\def\b{\begin{equation}}
\def\e{\end{equation}}
\def\n{\nonumber}
\def\L{\Lambda}
\def\D{\Delta}
\def\d{\delta}
\def\a{\alpha}
\def\o{\over}
\def\s{\sigma}
\def\t{\tilde}
\begin{document}
\draft
\title{UNITY OF FORCES AT THE PREON LEVEL WITH NEW GAUGE SYMMETRIES}
\author{M.K.Parida}
\address{Physics Department, North-eastern Hill University\\
 P.O.Box 21, Laitumkhrah, Shillong 793 003, India}
\date{February 3, 1997}
\maketitle
\begin{abstract}
In the context of a viable , supersymmetric, preon model, it has been shown 
by Babu and Pati that the unity of forces can well occur at the level of 
preons near the Planck scale. This preonic approach to unification is explored 
further in this note with the inclusion of threshold effects which arise due 
to spreading of  masses near the scale of supersymmetry ($M_S=1$ TeV) and the 
metacolor scale ($\L_M=10^{11}$ GeV). These effects, which were ignored in the 
earlier work, are found to have marked consequences in the running and 
unification of the relevant couplings leading to new possibilities for 
flavor-color as well as metacolor gauge symmetries. In particular, allowing 
for seemingly reasonable threshold effects, it is found that the metacolor 
gauge symmetry, $G_M$ is either $SU(6)_M$ or $SU(4)_M$ (rather than $SU(5)_M$) 
and the corresponding flavor-color gauge symmetry is  either $SU(2)_L\times 
U(1)_R\times SU(4)^C_{L+R}$ (for $G_M=SU(6)_M$) or even just the standard 
model symmetry $SU(2)_L\times U(1)_Y\times SU(3)_C$ (for $G_M=SU(6)_M$ or 
$SU(4)_M$). Prospects of other preonic gauge symmetries are also investigated.
\end{abstract} 
\pacs{12.60.Jv, 12.15.Ff, 12.60.Rc}
\narrowtext
\section{INTRODUCTION}
In the context of grand unification [1-3], it is known that, while the 
nonsupersymmetric minimal $SU(5)$ model [2] is excluded by proton decay 
searches [4] and by the recent LEP-data [5], the three coupling constants of 
the standard model approximately unify at a scale $M_X\approx 2\times 10^{16}$ 
GeV if one invokes supersymmetry e.g.~into minimal $SU(5)$ [6,7] or $SO(10)$. 
Despite this success, it seems to us, that neither of two schemes [$SU(5)$ or 
$SO(10)$] is likely to be a fundamental theory by itself because each scheme 
possesses a large number of arbitrary parameters associated with the Higgs 
sector; the corresponding Higgs exchange force in each case is thus not 
unified. Furthermore, neither scheme explains the origin of the three families 
and that of the diverse mass scales which span from 
the Planck mass $(\equiv M_{Pl})$ to 
$m_\nu$. These shortcomings are expected to be removed if one of the two 
schemes i.e., the minimal SUSY $SU(5)$ or the SUSY $SO(10)$ could emerge from 
superstring theory [8,9] which is such that it yields just the right spectrum 
of quarks, leptons, and Higgs bosons and just ``the right package" of Higgs 
parameters, thereby removing the unwanted arbitrariness. But, so far, this is 
far from being realized. An alternative possibility is that, instead of a 
grand unification symmetry, the minimal supersymmetric standard model with the 
``right package" of parameters might emerge directly from a superstring 
theory. 
In this case there is, however, the question of mismatch between the 
unification-scale $M_X$ obtained from, extrapolation of low-energy LEP-data, 
and the expected scale of string-unification which is nearly 20 times higher 
[10].
\par   
For these reasons, it has been suggested in an alternative approach that the 
unification of forces might occur as well at the level of constituents of 
quarks and leptons called the, ``preons" [11-15]. On the negative side, the 
preonic approach needs a few unproven, though not implausible, dynamical 
assumptions as regards the preferred direction of symmetry breaking and 
saturation of the composite spectrum [15-17]. On the positive side, it has the 
advantage that the model is far more economical in field content and 
especially in parameters than conventional grand unification models. The 
fundamental forces have a purely gauge origin, as in QCD, with no elementary 
Higgs bosons, and, therefore, no arbitrary parameters which are commonly 
associated with the Higgs-sector. Most important aspect of the model is that, 
utilising primarily the symmetries of the theory and the forbiddenness of SUSY 
breaking [18], in the absence of gravity, it provides simple explanation for 
the protection of composite quark-lepton masses [19]. The model seems capable 
of addressing successfully the origin of family unification and that of the 
diverse mass-scales [12], including interfamily mass-hierachy [14]. Finally it 
provides several testable predictions [12, 14-17].
\par
The question of unity of forces at preonic level was explored in a recent work 
by Babu and Pati [15], where it was shown that the unity occurs near the 
Planck scale ($\approx 10^{18}$ GeV), in accord with the LEP data, but with 
the flavor-color gauge symmetry $G_{fc}=SU(2)_L\times U(1)_R\times 
SU(4)_{L+R}^C$ and the metacolor gauge symmetry $G_M=SU(5)_M$. Considering 
that Planck-scale unification, as opposed to unity near $2\times 10^{16}$ GeV, 
goes better with the idea of string unification [8-10], we explore further, in 
this note, the preonic approach to unification, with the inclusion of 
threshold effects, which arise due to the spreading of masses near the scale 
of supersymmetry ($M_S\approx 1$ TeV ) as well as the metacolor scale
($\L_M\approx 10^{11}$ GeV). In particular, allowing for seemingly reasonable 
threshold effects, it is found that the unity of forces can well occur for 
certain desirable cases for which the metacolor gauge symmetry, $G_M$ is 
either $SU(6)_M$ or $SU(4)_M$ (rather than $SU(5)_M$) and the corresponding 
flavor-color gauge symmetry ($G_{fc}$) is either $SU(2)_L\times U(1)_R\times 
SU(4)^C_{L+R}$ (for $G_M=SU(6)_M$) or even just the standard model symmetry 
$SU(2)_L\times U(1)_Y\times SU(3)^C$ (for $G_M=SU(6)_M$ or $SU(4)_M$). These 
possibilities were disfavored in the earlier work because threshold effects 
had been ignored altogether. While estimating threshold effects at the 
supersymmetric and metacolor scales, we have used only bare masses excluding
wave-function-renormalisation corrections which have been shown by Shifman 
[20] to be cancelled by two-loop effects. We assure that such 
cancellation does not affect the results of this analysis and 
the threshold effects due to bare masses are enough to establish new gauge 
symmetries.
\par
An additional new result of this paper is the equality of one-loop 
$\beta$-function coefficients of $SU(2)_L$ and $SU(3)_C$ for $\mu>\L_M$ when 
these subgroups are embedded in $G_{fc}=SU(2)_L\times U(1)_Y\times SU(3)_C$, 
$SU(2)_L\times U(1)_R\times U(1)_{B-L}\times SU(3)_C$ or $SU(2)_L\times 
SU(2)_R\times U(1)_{B-L}\times SU(3)_C$ as long as the metacolor group is 
$G_M=SU(6)_M$. This implies one-loop partial unification of relevant gauge 
couplings above the metacolor scale.
\par 
This paper is organized in the following manner. In Sec.2 we present salient 
features of the scale-unifying preon model. The spectrum of composites near 
the electroweak and metacolor scales are given in Sec.3. Threshold effects due 
to composites are discussed in Sec.4. The equality of one-loop 
$\beta$-function coefficients for $SU(2)_L$, $SU(2)_R$ and $SU(3)_C$ using 
$G_M=SU(6)_M$ is proved in Sec.5 where the possibilities of different preonic 
gauge symmetries are also explored. The prospects of $SU(4)_M$ as metacolor 
gauge symmetry are explored in Sec.6. Results and conclusions of this work are 
summarised in Sec.7.
\section{SALIENT FEATURES OF THE SCALE UNIFYING PREON MODEL}
The effective Lagrangian below the Planck mass in the scale-unifying preon 
model [12] is defined to possess $N=1$ local supersymmetry and a gauge 
symmetry of the form $G_P=G_{fc}\times G_M$ where $G_M=SU(N)_M$ or $SO(N)_M$ 
denotes the metacolor gauge symmetry that generates the preon binding force. 
Although the underlying flavor-color gauge symmetry having preons in the 
fundamental representation has been suggested [12] to be 
$G_{fc}=SU(2)_L\times SU(2)_R\times SU(4)_{L+R}^C$ [1], any one of its 
subgroups could be a candidate for the effective flavor-color symmetry below 
the Planck scale [15],
\[G_{213}=SU(2)_L\times U(1)_Y\times SU(3)_C,\n\]
\[G_{2113}=SU(2)_L\times U(1)_R\times U(1)_{B-L}\times SU(3)_C,\n\]
\[G_{2213}=SU(2)_L\times SU(2)_R\times U(1)_{B-L}\times SU(3)_C,\n\]
\[G_{214}=SU(2)_L\times U(1)_Y\times SU(4)^C_{L+R},\n\]
\b G_{224}=SU(2)_L\times SU(2)_R\times SU(4)^C_{L+R}.\e 
Here $G_{2213}$ and $G_{224}$ are assumed to possess left-right discrete 
symmetry ($=$Parity) leading to $g_{2L}(\mu )=g_{2R}(\mu )$ for $\mu\agt\L_M$. 
The gauge symmetry $G_P$ operates on a set of preonic constituents consisting 
off six positive and six negative chiral superfields while each of these 
transforms as the fundamental representation $N$ of $G_M=SU(N)_M$,
\[\Phi^a_\pm=(\phi^a_{L,R},\ \ \psi^a_{L,R},\ \ F^a_{L,R}),\ \  
a=(x,y,r.y.b.l)\n\]
Here(x,y) denote the two basic flavor attributes (u,d) and, (r,y,b,l), the 
basic color attributes of a quark lepton family [1]. Thus, $\Phi^{X,Y}_+$ and 
$\Phi^{X,Y}_-$ transform as doublets under $SU(2)_L$ and $SU(2)_R$, 
respectively, while both $\Phi^{r,y,b,l}_+$ and $\Phi^{r,y,b,l}_-$ transform 
as quartets under $SU(4)^C_{L+R}$. The effective Lagrangian of this 
interaction turns out to possess only gauge and gravitational interactions 
and, as a result, involves only three or four coupling constants of the 
gauge symmetry $G_{fc}\times G_M$.
\par
The model has profound interpretation of hierarchy of mass scales as follows 
[12]. Corresponding to an input value of the metacolor coupling 
$\t\a_M=1/20-1/30$ at $M_{Pl}/10$, the asymtotically free metacolor force 
generated by $SU(N)_M$ becomes strong at scale $\L_M\approx 10^{11}$ GeV for 
$N=4-6$. Thus, one small number ($\L_M/M_{Pl})\sim 10^{-8}$ arises naturally 
through renormalization group equations(RGEs) due to small logarithmic growth 
of $\t\alpha_M$ and its perturbative input value at $M_{Pl}/10$. The remaining 
small scales arise primarily due to the Witten index theorem [18], which would 
forbid a dynamical breaking of SUSY, if there was no gravity. Noting that both 
the metagaugino condensate $\langle\vec\lambda.\vec\lambda\rangle$ and the 
preonic condensate $\langle\bar\psi^a\psi^a\rangle$ break SUSY (for massless 
preons), they must both need the collaboration between the metacolor force and               
gravity to form. Assuming that they do form, one can argue plausibly that they 
must each be damped by a factor $\L_M/M_{Pl}$ [20]. Since $\langle\bar\psi^a
\psi^a\rangle$ breaks not only SUSY but also $SU(2)_L\times U(1)_Y$  for 
$a=x,y$, one obtains SUSY breaking mass splittings $\d m_S\sim\L_M(\L_M
/M_{Pl})\sim$ 1 TeV and $M_W\sim (1/10)\L_M(\L_M/M_{Pl})\approx 100$ GeV. The 
symmetry of the fermion mass matrix involving three chiral families, 
$q^i_{L,R}$, and two vectorlike families, $Q_{L,R}$ and $Q'_{L,R}$, where the 
chiral families acquire mass almost through their mixings with vectorlike 
families by seesaw mechanism [12], explains the interfamily hierarchy 
$(m)_{u,d,e}\ll (m)_{c,s,\mu}\ll (m)_{t,b,\tau}$ with $m_{u,d,e}\sim O(1)$ MeV 
and $m_t\sim M_W\sim 100$ GeV [14]. Finally a double seesaw mechanism with 
$m(\nu_R^i)\sim\L_M\sim 10^{11}$ GeV and $m(\nu)_{\rm Dirac}\sim\L_M(\L_M
/M_{Pl})$ yields $m(\nu^i_L)<10^{-3}M_{Pl}(\L_M/M_{Pl})^3\sim 10^{-27}M_{Pl}$. 
In this way the model provides, remarkably enough, a common origin of all the 
diverse scales from $M_{Pl}$ to $m_{\nu}$ [12].                                            
\par
Owing to the fermion-boson pairing in SUSY, the model also turns out to 
provide a good reason for family replication and (subject to the saturation at 
the level of minimum dimensional composite operators) for having just three 
chiral families $q^i_{L,R}$ [13]. It also predicts two complete vectorlike 
families $Q_{L,R}=(U,D,N,E)_{L,R}$ and $Q'_{L,R}=(U',D',N',E')_{L,R}$ with 
masses of order 1 TeV where $Q_{L,R}$ couple vectorially to $W_L$'s and 
$Q'_{L,R}$ to $W_R$'s. The masses of the superpartners of all fermions are 
predicted to be $(0.5-2)$ TeV.                                                                   
\par
The model presumes that the preonic condensate $\D_R$, transforming under 
$G_{224}$ as $(1,3_R,10^{*C})$, is formed and its neutral component acquires 
VEV, $\langle\D^0_R\rangle\simeq\L_M\simeq 10^{11}$ GeV which preserves SUSY 
but breaks $G_{224}$ and its subgroups to $G_{213}$. Finally the condensate 
$\langle\bar\psi^a\psi^a\rangle$, for $a=x,y$, breaks SUSY as well as the 
electroweak gauge symmetry, $SU(2)_L\times U(1)_Y$. As a result, the model 
leads to many consequences common with a two-step breaking of $SO(10)$. 
Subject to left-right symmetry, the effective Lagrangian has three gauge 
couplings with $G_{224}\times SU(N)_M$, four with $G_{2213}\times SU(N)_M$, 
$G_{214}\times SU(N)_M$, and $G_{213}\times SU(N)_M$, but five with 
$G_{2113}\times SU(N)_M$. Furthermore, if the gauge symmetry $G_P$ and the 
associated preon content specified above arise from an underlying superstring 
theory, in particular, through a four-dimensional construction [9] with $k=1$ 
Kac-Moody algebra, the few gauge coupling constants of the model would be 
equal to one coupling at the string unification scale $M_U\sim 10^{18}$ GeV 
(barring string threshold effects) [10]. It is this posibility of 
gauge-coupling unification at the preon level, with $G_M=SU(6)_M$ and 
$SU(4)_M$, which is explored in this paper including threshold effects at 
$M_{\rm SUSY}$ and $\L_M$.
\par
As it is well known that the flavor symmetry near $\mu=100$ GeV is given by 
the standard gauge symmetry $G_{213}$ with quarks and leptons in the 
fundamental representation, and that at low energies is $U(1)_{em}\times 
SU(3)_C$, it might appear that the five flavor-color symmetries given in 
eq.(1) have been arbitrarily chosen for the preonic effective Lagrangian. But 
realizing that the two important ingredients in the model [12] are left-right 
symmetry and $SU(4)$-color[1], the flavor-color symmetry $G_{224}$ has been 
suggested as the natural gauge symmetry near the Planck scale in the presence 
of $G_M=SU(N)_M$. Thus, below $\mu=M_{Pl},G_{224}$ itself or any of its four 
subgrups given in eq.(1) could be natural choices for the preonic effective 
Lagrangian. However, in addition to the assumed saturation of minimum 
dimensional operator and the composite spectrum, the model has arbitrariness 
in that it does not specify a unique direction of symmetry breaking. This 
latter feature is also commom to  the usual SUSY $SO(10)$ with more than one 
choices for intermediate gauge symmetries. But, nevertheless, the preons 
combine to form quarks and leptons, and Higgs scalars near $\mu=\L_M$ due to 
the strong metacolor binding force and every other $G_{fc}$, except $G_{213}$, 
undergoes spontaneous symmetry breaking leadig to the standard gauge symmetry. 
In addition to the three standard families of quarks and leptons, the new 
vectorial fermions are predicted to have masses near 1 TeV which can be 
testified by accelerator experiments [13-17]. The right handed neutrinos 
aquire masses near $\L_M$ and contribute to the see-saw mechanism. 
\section{SPECTRUM OF COMPOSITES NEAR ELECTROWEAK AND METACOLOR SCALES}
In this section we discuss briefly the spectrum of massive particles near the 
electrowek scale ($M_Z$) and the metacolor scale $(\L_M\simeq 10^{11}$ GeV).             
In the scale-unifying preon model the left-handed and the right-handed chiral 
fermions in each of the three families transform as $(2_L,1,4^{*C})$ and 
$(1,2_R,4^{*C})$, respectively, under $SU(2)_L\times SU(2)_R\times 
SU(4)_{L+R}^C$ [1]. The two vector-like families $Q_{L,R}$ and $Q'_{L,R}$ 
transform  as $(2_L,1,4^{*C})$ and $(1,2_R,4^{*C})$, respectively. The members 
of five families predicted by the scale-unifying preon model [12-13] are 
denoted by 
\[q^e_{L,R}=(u,d,\nu,e )_{L,R}\n\]
\[q^\mu_{L,R}=(c,s,\nu_\mu,\mu)_{L,R}\n\]
\[q^\tau_{L,R}=(t,b,\nu_\tau,\tau)_{L,R}\n\]
\[Q_{L,R}=(U,D,N,E)_{L,R}\n\]
\b Q'_{L,R}=(U',D',N',E')_{L,R}\n\e
The spectrum of light and heavy particles including matter multiplets near the 
electroweak and the metacolor scales and their quantum numbers under the gauge 
groups $G_{224}$ and $G_{213}$ are summarized in Table I. In order to compute 
threshold effects, we present, in Table II, assumed but plausible values of 
masses for the Higgs scalars and different members of vectorial families along 
with the current experimental value for $m_t$, including their contributions 
to one-loop $\beta$-function coefficients\footnote{Two vector-like families 
have the quantum numbers of a $16+\overline{16}$ of $SO(10)$. Thus, their 
contributions to $\beta$-functions are the same as those of two standard 
chiral families.}. The corresponding values for all the superpartners of the 
standard chiral families, gauginos and the Higgsinos are given in Table III. 
For the sake of simplicity, all the superpartners of two vector-like families 
are assumed to be degenerate at the scale $M_S=1.5$ TeV above which SUSY is 
assumed to be restored\footnote{Changing the superpartner scale from $M_S=1.5$ 
TeV, used in this analysis, to $M_S=1$ TeV would increase the value of the              
strong interaction coupling by less than a few percent without any significant 
change on the results and conclusions.}. As usual, there are two Higgs 
doublets, $u$-type and $d$-type, near the electroweak scale contained in the 
$G_{224}$-submultiplet $\phi (2,2,1)$ which is a two-body  condensate made out 
of the preons.
\par
As noted in Sec.2, it is essential that the four-body preonic condensate, 
$\D_R(1,3,10^{*C})$, is formed with mass near $\L_M$ to drive the seesaw 
mechanism resulting in small values of left-handed Majorana neutrino mass. The 
underlying left-right symmetry of the effective Lagrangian then requires the 
formation of the corresponding composite $\D_L(3,1,10^C)$. In fact presevation 
of SUSY down to 1 TeV scale, especially through the D-term, may require an 
additional pair, $\bar\D_L+\bar\D_R$, having masses same as their counterparts 
in the first pair. In what follows we will drop the distinctions between 
$\D_i$ and $\bar\D_i$ $(i=L,R)$ as both have identical contributions to 
$\beta$-functions. Thus, two sets of $\D_L$ and $\D_R$ are the minimal 
requirements of the scale-unifying preon model. Before $\D^O_R$ acquires VEV 
$\simeq\L_M\simeq 10^{11}$ GeV, the massess of $\D_L(\bar\D_L)$ and 
$\D_R(\bar\D_R)$ are identical. But the VEV splits them leading to their mass 
ratio which could be as large as 3.                    
\par
In specific cases, we will also assume the formation of composite 
Higgs-supermultiplets of the type $\s (1,1,15)$ and $\xi (2,2,15)$ under 
$G_{224}$ as optional choices. It is to be noted that while the field $\s 
(1,1,15)$ is a two-body composite, $\xi (2,2,15)$ is a four body composite. 
Since the masses of these composites are not constrained by the VEV of $\D_R$, 
they are allowed to vary over a wider range around $\L_M$ as compared to the            
masses of $\D_L$ and $\D_R$. It can be argued that more than one set of $\s$ 
and $\xi$ fields are allowed to form but we will confine to at most two such 
sets with masses $(1-7)\L_M$ or $(1/7-1)\L_M$ as the case may be. 
All the masses used for estimation of threshold effects near the metacolor
scale as well as the supersymmetry breaking scale are bare masses devoid of 
wave-function-renormalisation which is shown to be cancelled out by two-loop 
effects [20]. We assure that the threshold effects due to bare masses are
enough to establish new gauge symmetries and the observed cancellation [20]
does not affect the results of this paper.
In Tables IV and V we present the superheavy-particle spectrum near the 
metacolor scale with their respective quantum numbers under $G_{224}$ and 
$G_{213}$.                                                 
\section{THRESHOLD EFFECTS AT LOWER AND INTERMEDIATE SCALES}
In this section we discuss renormalization group equations (RGEs) [21] for 
gauge couplings in the scale-unifying preon model using the gauge symmetry 
$SU(2)_L\times U(1)_Y\times SU(3)^C(=G_{213})$ for the composite quarks, 
leptons and Higgs scalars and their superpartners between $M_Z$ and $\L_M$. At 
first the gauge couplings of $G_{213}$ are evolved from $M_Z$ to $\L_M$ 
assuming SUSY breaking scale to be $M_S=1.5$ TeV and including 
the threshold effects at $M_Z$ and $M_S$ through the matching functions 
$\Delta_i^{(Z)}$, and $\Delta_i^{(S)}$, respectively [22-23]. The RGEs for the 
three gauge couplings of $G_{213}\ \ (i=1,2,3)$ are         
\[{1\o\a_i(M_Z)}={1\o\a_i(\L_M)}+{b_i\o 2\pi}\ln{M_S\o M_Z}+
{b'_i\o 2\pi}\ln{\L_M\o M_S}-\D_i^{(L)}\n\]
\b\D_i^{(L)}=\D_i^{(Z)}+\D_i^{(S)}\e
where we have neglected the two-loop effects. Threshold effects at $\L_M$ have 
been included in the second part of this section. The L.H.S. of (3) is 
extracted using the CERN-LEP data and improved determination of the 
finestructure constant at $M_Z=91.18$ GeV  [5],            
\[\sin^2\theta_W(M_Z)=0.2316,\n\]                                                       
\[\a^{-1}(M_Z)=127.9\pm 0.2,\n\]                                                     
\b\a_S(M_Z)=0.118\pm 0.007,\e
leading to the following values of couplings\footnote{The value of 
$\sin^2\theta_W(M_Z)=0.2316$ is cosistent with a heavy top ($m_t=175$ GeV). We 
ignore negligible threshold effects due to the top-quark mass on electroweak 
gauge couplings.} of $G_{213}$ at $M_Z$,               
\[\a_1^{-1}(M_Z)=58.96,\n\]
\[\a_2^{-1}(M_Z)=29.62,\n\]                                                          
\b\a_3^{-1}(M_Z)=8.33\pm 0.63.\e                                  
The matching functions $\D_i^{(Z)}$ include threshold effects due to the top 
quark coupling to the photon, the electroweak gauge bosons, gluons, and its 
Yukawa coupling to the Higgs scalars [23]. The contributions due to the two 
Higgs doublets, the additional fermions of two vectorlike families ($Q,Q'$) 
and all superpartners, having specific values of masses within a given range 
but below $M_S$ are included in $\D_i^{(S)}$. The one loop coefficients $b_i$ 
in (3) are computed using three generations of fermions $(n_g=3)$ and 
excluding the contributions of the Higgs doublets $(n_H=0)$ and vectorlike            
families. Since the contributions of the Higgs scalars and vectorlike families 
are included in $\D_i^{(S)}$ incorporating the specific assumptions on their 
masses, the approach adopted here is equivalent to the conventional approach 
as contribution due to every particle to the gauge-coupling evolution is 
accounted for,             
\b b_i=-{11\o 3}t_2(V)+{2\o 3}\sum t_2(F)+{1\o 3}\sum t_2(S)\e
where $t_2(V)$, $t_2(F)$, and $t_2(S)$ denote the contributions of gauge 
bosons, fermions and Higgs-scalars, respectively. For an $SU(n)$ group with 
matter in the fundamental representation and gauge bosons in the adjoint,
\[t_2(F)=t_2(S)=1/2,\ \ t_2(V)=n\n\]                                                 
whereas $t_2(V)=0$ for any $U(1)$ group. With supersymmetry, (6) gives,
\b b'_i=-3t_2(V)+\sum t_2(F)+\sum t_2(S)\e                
In the region I where $\mu=M_Z$ to $M_S=1.5$ TeV, we evaluate the coefficients 
by including the contributions of gauge bosons and three standard fermion 
generations (as all other contributions in this region are included in 
$\D_i^{(S)}$),
\[b_3=-{11\o 3}\times 3+{4\o 3}\times 3=-7\n\]
\[b_2=-{11\o 3}\times 2+{4\o 3}\times 3=-{10\o 3}\n\]
\b b_1={4\o 3}\times 3=4\e
In the region II where $\mu=M_S$ to $\L_M$, the spectrum of particles consists 
off the gauge bosons of $G_{213}$, the three normal families of fermions 
$(n_g=3)$, two additional vectorlike families corresponding to $n'_g=n_g+2$, 
the two Higgs doublets and superpartners of these particles such that SUSY is 
restored for $\mu>M_S=1.5$ TeV. Using (7) we evaluate,                                  
\[b'_3=-3\times 3+2n'_g=1\n\]
\[b'_2=-3\times 2+2n'_g+2\times{1\o 2}=5\n\]
\b b'_1=2n'_g+{2\o 5}={53\o 5}\e
Now we discuss explicitly how threshold effects at the boundaries $M_S$ and 
$M_Z$ are evaluated.                                                      
\subsection{Threshold Effects at Lower Scales}
The top-quark threshold contribution which is the same in SUSY and nonSUSY 
standard model has been discussed in ref.[23]. Since the value of $\sin^2
\theta_W$ in (4) is consistent with the experimental value of top quark mass, 
$m_t=175$ GeV, we ignore negligible electroweak threshold corrections due to 
the heavy top but include those on $\a_3^{-1}(M_Z)$ and Yukawa coupling 
corrections. The coupling of the top-quark to gluons gives 
rise to
\b\D^{top}_3={1\o 3\pi}\ln{m_t\o M_Z}=.07\e
The top-quark mass $m_t=175$ GeV is consistent with its Higgs-Yukawa coupling 
$h_t\simeq 1$ leading to the threshold corrections at two-loop level,
\b\D^{Yuk}_i={h^2_t\o 32\pi^3}\left(b_i^{top}\ln{M_S\o 174{\rm GeV}}+
b^{'top}_i\ln{\L_M\o M_S}\right)\e
where $b^{top}_i=(17/10,3/2,2)$ for $i=1,2,3$ in the standard model and            
$b_i^{'top}=(26/5,6,4)$ in the MSSM. Using $M_S=1.5$ TeV, and $\L_M=10^{11}$ 
GeV gives,                                                                        
\b\D_1^{Yuk}=0.10,\ \ \D_2^{Yuk}=0.12,\ \ \D_3^{Yuk}=0.08\e
Adding the contributions in (10) and (12) yields,                             
\b\D_1^{(Z)}=0.10,\ \ \D_2^{(Z)}=0.12,\ \ \D_3^{(Z)}=0.15\e
It is clear that the corrections are smaller and unlikely to affect our 
analysis unless the Yukawa couplings of heavy families are much 
larger\footnote{Since the masses of vector-like families occur as off-diagonal              
elements, they receive no contributions from the Yukawa couplings of the two 
Higgs doublets of the standard SUSY model. Hence their Yukawa contributions to 
threshold effects is likely to be smaller.} i.e., $h_{Q,Q'}=3-5$.                                            
\par
Threshold effects at $M_S$ due to masses below it are computed explicitly 
using the second and the third terms in (6) depending upon the nature of the 
particle ``$\a$",
\b\D_i^{(S)}=\sum_\a{b_i^\a\o 2\pi}\ln{M_\a\o M_S}\e
The values of $b_i^\a$ and the masses  $M_\a$ used in this analysis are given 
in Tables II and III for each particle which lead to,               
\b\D_1^{(S)}=^{-1.8}_{-1.0},\ \ \D_2^{(S)}=^{-2.3}_{-0.9},\ \ 
\D_3^{(S)}=^{-1.8}_{-1.1}\e
Combining (13) and (15) gives the following threshold corrections at lower 
scales,                                                              
\b\D_1^{(L)}=^{-1.70}_{-0.90},\ \ \D_2^{(L)}=^{-2.20}_{-0.80},\ \ 
\D_3^{(L)}=^{-1.65}_{-0.95}\e
In (15) and (16) the upper and lower entries are due to lowest and the highest 
values of $M\a$  given in Tables II and III. The evolution of the gauge 
couplings upto $\mu=\L_M$ includindg threshold effects at $M_Z$ and $M_S$, 
but excluding those at $\L_M$ yields,            
\[\a^{-1}_1(\L_M)=26.6\ \ (25.6)\n\]
\[\a^{-1}_2(\L_M)=16.0\ \ (15.67)\n\]
\b\a^{-1}_3(\L_M)=7.6\pm 0.6\ \ (6.9\pm 0.6)\e
where the quantities inside (outside) the parenthesis in (17) are due to the 
lowest (highest) values of $\D_i^{(L)}$ in (16). The gauge couplings at the 
metacolor scale are then obtained as,                       
\[g_1(\L_M)=0.685\ \ (0.700)\n\]                                                   
\[g_2(\L_M)=0.833\ \ (0.894)\n\]
\b g_3(\L_M)=1.28\pm 0.05\ \ (1.35\pm 0.06)\e
\subsection{Threshold Effects at the Metacolor Scale}
As explained in Secs.2 and 3, we will use two sets of the Higgs superfields 
$\D_L(3,1,10^C)$ and $\D_R(1,3,10^{*C})$ in all cases and two sets of $\xi 
(2,2,15)$ and  $\s (1,1,15)$, wherever necessary. Denoting $\a'_i(\L_M)$ for 
the gauge couplings of $G_{213}$ at $\L_M$ including threshold  effects 
through the matching functions $\d_i$, they are related to $\a_i(\L_M)$ of (5) 
and (17) as,                                     
\b{1\o\a_i(\L_M)}={1\o\a'_i(\L_M)}-\d_i\e
In addition to the superheavy-particle-threshold effects, $\d_i$ may have a 
very small correction due to conversion from $\overline{DR}$ to 
$\overline{MS}$ scheme [23] in the relevant cases\footnote{The term 
$\D_i^{CNV}=-C_2(G_i)/12\pi$ where $C_2(G_i)=N$ for $SU(N)$, but $C_2(G_i)=0$ 
for $U(1)$, appears from the necessity to use $\overline{DR}$ scheme.}. The 
matching functions $\d_i$ are evaluated by one-loop approximation as,                                   
\b\d_i=\sum_\rho{b_i^\rho\o 2\pi}\ln{M_\rho\o\L_M}=
\sum_\rho{b_i^\rho\o 2\pi}\eta_\rho\e
where $\rho$ runs over all the submultiplets of a $G_{224}$-multiplet and we             
have used the notation $\eta_\rho=\ln(M_\rho /\L_M)$. The decomposition of 
each $G_{224}$ representation under $G_{213}$ and the contribution to the 
one-loop $\beta$-function coefficient $(=b_i^\rho)$ are 
presented in Tables IV and V. Since the exact values of the 
masses of the submultiplets are not predicted by the model, we make the 
simplifying assumtion that all the submultiplets belonging to the same 
$G_{224}$-multiplet have a degenerate bare mass [20]. Including all possible 
contributions due to the $G_{213}$-representations of Tables IV and V we 
obtain,                    
\[\d_1={1\o 10\pi}\left( 77\eta_\xi+18\eta_{\D_L}+
78\eta_{\D_R}+8\eta_\s\right)\n\]
\[\d_2={1\o 2\pi}\left( 15\eta_\xi+20\eta_{\D_L}\right)\n\]
\b\d_3={1\o 2\pi}\left( 16\eta_\xi+9\eta_{\D_L}+9\eta_{\D_R}+
4\eta_\s\right)\e
There are slight variations from (21) in specific cases depending upon the 
preonic gauge symmetry given in (1). In the case of $G_{fc}=G_{2213}$, certain 
components of $\D_R(1,3.10^{*C})$ are absorbed as longitudinal modes of 
$SU(2)_R$ gauge bosons leading to                  
\b\d_1={1\o 10\pi}\left( 77\eta_\xi+18\eta_{\D_L}+75\eta_{\D_R}+
8\eta_\s\right)\e  
but the expressions  for $\d_2$ and $\d_3$ are the same as in (21). Similarly 
when $G_{fc}=G_{224}$, the submultiplet having the $G_{213}$-quatum numbers 
$(1,2/3,\overline{3})$ is absorbed as longitudinal mode of massive $SU(4)_C$ 
gauge bosons and does not contribute to $\d_1$  and $\d_3$,            
\[\d_1={1\o 10\pi}\left( 77\eta_\xi+18\eta_{\D_L}+71\eta_{\D_R}+
8\eta_\s\right)\n\]
\b\d_3={1\o 2\pi}\left( 16\eta_\xi+9\eta_{\D_L}+{17\o 2}\eta_{\D_R}+
4\eta_\s\right)-{1\o 4\pi}\e  
where the term $-(4\pi)^{-1}$ arises due to conversion from $\overline{DR}$ to 
$\overline{MS}$ scheme [23]. The exprerssion for $\d_2$ in this case is the 
same as in (21). In the case of $G_{fc}=G_{214}$,                                              
\b\d_1={1\o 10\pi}\left( 77\eta_\xi+18\eta_{\D_L}+
74\eta_{\D_R}+8\eta_\s\right)\e  
but the expressions for  $\d_2(\d_3)$ are given by (21)((23)).                    
\section{PREONIC GAUGE SYMMETRIES AND UNIFICATION OF GAUGE COUPLINGS}
In this section we explore possible gauge symmetries of the preonic effective 
Lagrangian that operates from  $\mu=\L_M\simeq 10^{11}$ GeV to $M_U(=M_{Pl}/
10=10^{18}$ GeV). In ref.[15] it has been successfully demonstrated that unity 
of fundamental forces occurs with preons as fermion representations of the 
gauge group $G_P=SU(2)_L\times U(1)_R\times SU(4)^C_{L+R}\times SU(5)_M$. In 
this section we confine to prospects of $SU(6)_M$. In what follows we searh 
for converging solutions to gauge couplings as we approach towards $M_{Pl}$ 
We prefer approximate to exact unification of the gauge couplings as 
the gravitaional effects are to make substantial contributions which might 
compensate for the remaining small differences.
\par
The RGEs for the gauge couplings ($\t\a_i(\mu)=\t g^2_i(\mu)/4\pi$) of the                
preonic effective lagrangian for $\mu=\L_M$ to $M_U$ can be written at one 
loop level as [21, 23],                                                       
\b{1\o\t\a_i(\L_M)}={1\o\t\a_i(\mu)}+{b''_i\o 2\pi}\ln{\mu\o\L_M}\e
where $b''_i$ is the one-loop coefficient of the $\beta$-function with 
preons in the fundamental representation, which are separately evaluated in 
each case. For computation of threshold effects, while a mass ratio 
$\rho=M_{\D_L}/M_{\D_R}=2-3$ could be considered natural, we also keep an open 
mind to explore unification possibilities with such values of inverse mass 
ratio. We adopt the strategy of examining approximate unification starting 
from smaller values of $\D_L-\D_R$ mass difference within 10\% to 20\% and 
then increasing the mass difference corresponding to higher values of $\rho$. 
When we find that approximate unification is not achievable with the minimal 
two sets of $\D_L$ and $\D_R$ fields, we introduce threshold effects due to 
the two optional sets of fields $\xi (2,2,15)$ and $\s (1,1,15)$. We report 
our investigations in different cases.
\subsection{$G_P=SU(2)_L\times U(1)_R\times SU(4)_{L+R}^C\times SU(6)_M$}                          
Corresponding to $G_P=G_{214}\times SU(6)_M$, $i=1R,2L,4C,$ and 6 in (25) and 
the one-loop coefficients are,                                           
\[b''_{1R}=3,\ \ b''_{2L}=-3,\ \ b''_{4C}=-6,\ \ b''_6=-12\n\]
The matching conditions between the gauge couplings of elementary preons 
($\t g_i(\mu)$) and composite fields ($g_i(\mu)$) at $\mu=\L_M$ are written 
as            
\bm\b\a^{-1}_2(\L_M)+\d_2=\t\a^{-1}_{2L}(\L_M)\e
\b\a^{-1}_3(\L_M)+\d_3=\t\a^{-1}_{4C}(\L_M)\e
\b\a^{-1}_1(\L_M)+\d_1={3\o 5}\t\a^{-1}_{1R}(\L_M)+
{2\o 5}\t\a^{-1}_{4C}(\L_M)\e\em
where the L.H.S. in (26a)-(26c) are $\a^{'-1}_i(\mu=\L_M)$ $(i=1,2,3)$ of 
(19). Using (26b) and (17) in (26c) gives                                          
\b{3\o 5}\t\a^{-1}_{1R}(\L_M)=\d_1-{2\o 5}\d_3+{2\o 5}\d_3+23.66\pm 0.38\e
which yields $\t\a^{-1}_{1R}(\L_M)$ once $\d_1$ and $\d_3$ are specified. 
Excluding threshold effects at $\mu=\L_M(\d_i=0)$ and extrapolating the gauge            
couplings to $\mu=10^{18}$ GeV gives $\t\a^{-1}_{2L}(M_U)=23.8$ and 
$\t\a^{-1}_{4C}(M_U)=23.0$. This implies that when threshold effects are 
included, $\d_2=\d_3=7$ for approximate unification of gauge couplings with 
$SU(6)_M$ corresponding to $\t\a^{-1}_6(M_U)=30$ provided the matching 
condition (27) is satisfied with suitable values of $\d_1$ and $\t\a^{-1}_{1R}
(\L_M)$. It is found that these threshold corrections are significantly less                 
compared to other models with $SU(6)_M$ investigated in this paper.             
\par
To see how unification is achieved we start with $\d_2=8$ and $\d_3=7$. Then 
using (21) and (23) and setting  $\eta_\s=\eta_\xi=0$ we obtain              
\b\eta_{\D_L}=2.5,\ \ \eta_{\D_R}=2.4\e                 
In the presence of only the minimal number of two sets of fields, $\D_L+\D_R$ 
and $\bar\D_L+\bar\D_R$, as mentioned in Sec.3, eq.(28) implies\footnote{In 
our notation $\D_{L,R}\equiv\D^1_{L,R}$ and $\bar\D_{L,R}\equiv\D^2_{L,R}$ for 
the minimal two sets of fields needed from considerations of left-right 
symmetry, spontaneous breaking of gauge symmetry or generation of R.H. 
Majorana neutrino mass and preservation of SUSY down to the TeV scale. (see 
also Tables I and V)}                  
\[M_{\D_L}=M_{\bar\D_L}=3.5\times 10^{11}{\rm GeV},\] 
\b M_{\D_R}=M_{\bar\D_R}=3.2\times 10^{11}{\rm GeV}\e     
which differ by only 10\%. It is to be noted that these are bare masses 
including splitting due to VEV of $\D_R^0$, since the wave functions 
renormalisation effects have been shown to be cancelled by two-loop 
contributions [20]. The values of $\t\a^{-1}_{1R}(\L_M)$ are obtained from 
(27) as $\d_1$ is determined using (28) and  $\eta_\s=\eta_\xi=0$ in (23). 
Then $\t\a^{-1}_{1R}(M_U)$ is known through its RGE. With  $\t\a_6^{-1}(10^
{18})$ GeV $=29$, the gauge couplings at three different scales, $\mu=10^{11}$ 
GeV, $10^{18}$ GeV, and $10^{19}$ GeV are presented in Table VI. It is clear 
that the least difference between the gauge couplings, which is 2\% to 3\%, 
occurs near $10^{19}$ GeV i.e., the unification appears to occur at a scale 
one order higher than expected. The evolution of gauge couplings in this model 
has been shown in Fig.1. Even though the mass difference between
$\D_L(\bar\D_L)$ and $\D_R(\bar\D_R)$ is small, the strong interaction 
coupling $(g_{3C}(\L_M))$ of composite fields and the $SU(4)^C_{L+R}$ coupling 
of preons $(\t g_{4C}(\L_M))$ exhibit nearly 35\% difference due to threshold 
effect at $\mu=\L_M$. Similarly $g_{2L}(\L_M)$ and $\t g_{2L}(\L_M)$ show 
nearly 20\% difference. These are due to the fact that the individual masses 
of the two sets of fields given in (29) deviate from $\L_M$ by a factor 3.2 to 
3.5 which contribute to such significant threshold corrections. The remaining 
small differences among the gauge couplings at $\mu=10^{19}$ GeV are expected 
to be compensated by gravitational effects .
\begin{figure}
\begin{center}
\leavevmode
\hbox{\epsfxsize=4in
\epsfysize=4in
\epsffile{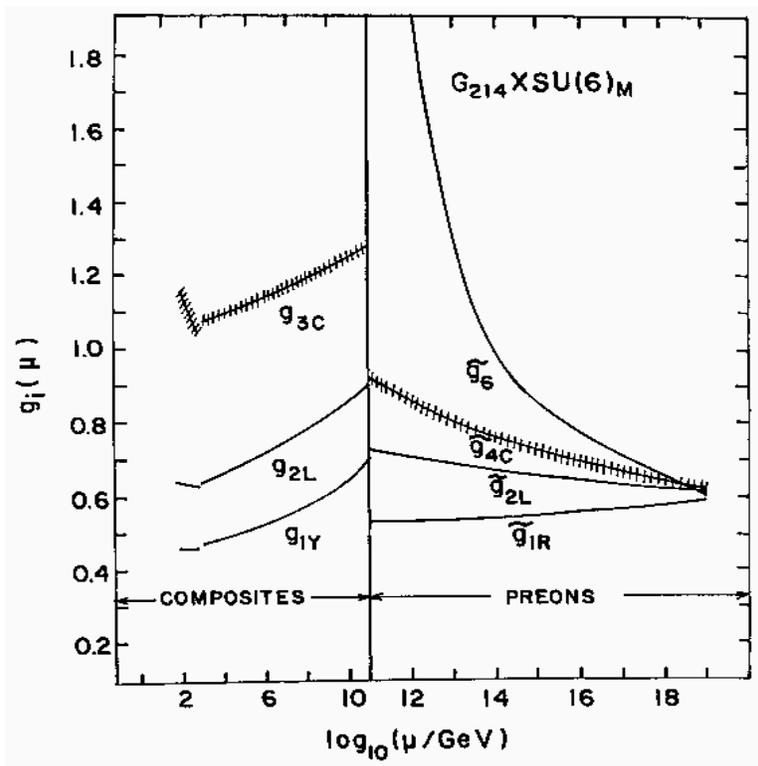}}
\end{center}
\caption{Unification of gauge couplings and their evolutions from
$M_{Pl}=10^{19}$ Gev to $M_Z$ including threshold effects at lower              
and intermediate (metacolor) scales for the preonic gauge            
symmetry, $SU(2)_L\times U(1)_R\times SU(4)^C_{L+R}\times SU(6)_M$, with 
minimal two sets $\D_L$ and $\D_R$ fields and 10\% mass difference between 
them.}
\end{figure}
\subsection{$G_P=SU(2)_L\times SU(2)_R\times U(1)_{B-L}\times SU(3)_C\times 
SU(6)_M$}
In this case we, assume the gauge group to possess the left-right discrete 
symmetry starting from $\mu=\L_M$ to $M_{Pl}$ with $\t g_{2L}(\mu)=            
\t g_{2R}(\mu)$. Denoting $i=1_{BL},2_L,2_R,3_C$ and 6 in (25), the one loop             
coefficients are,                                                             
\[b''_{BL}=6,\ \ b''_{2L}=b''_{2R}=-6+{6\o 2}=-3\n\]
\b b''_{3C}=-9+6=-3,\ \ b''_6=-3\times 6+6=-12\e
The equality of the coefficients $b''_{2L}=b''_{2R}=b''_{3C}$ signify           
unification of $SU(2)_L$ and $SU(3)_C$ gauge couplings from $\mu=\L_M$ to               
$M_{Pl}$ at one-loop level when preons are in the fundamental            
representation and the metacolor symmetry is $SU(6)_M$. This is a common 
feature for $G_{fc}=G_{213}$, $G_{2113}$, and $G_{2213}$ when $G_M=SU(6)_M$ as 
can be seen in the following manner:
\par
Suppose that $G=SU(N)_M$ for all the three types of $G_{fc}$. Then the 
one-loop coefficients for $SU(2)_L$ and $SU(3)_C$ are             
\b b''_{2L}=-6+{N\o 2},\ \ b''_{3C}=-9+N\e
The one-loop unification for all values of $\mu$ starting from $\mu=\L_M$ to 
$\mu=M_U$ is guarranted by the RGEs provided,                            
\b b''_{2L}=b''_{3C}=b''_i\e 
with                                                                           
\b{1\o\t\a_i(\mu)}={1\o\t\a_i(M_U)}+{b''_i\o 2\pi}\ln{M_U\o\mu},\ \ 
i=2L,3C\e
since $\t\a_{2L}(M_U)=\t\a_{3C}(M_U)$. But the equations (31) - (33) imply               
\b N=6\e
proving that the metacolor gauge group is $SU(6)_M$ to achieve such one-loop 
unification from $\mu=\L_M$ to $M_U$.
\par
The matching conditions with $G_{fc}=G_{2213}$ at $\mu=\L_M$ are                  
\[\a^{-1}_2(\L_M)+\d_2=\t\a^{-1}_{2L}(\L_M)=\t\a^{-1}_{2R}(\L_M)\n\]
\[\a^{-1}_3(\L_M)+\d_3=\t\a^{-1}_3(\L_M)\n\]
\b\a^{-1}_1(\L_M)+\d_1={3\o 5}\t\a^{-1}_{2R}(\L_M)+
{2\o 5}\t\a^{-1}_{BL}(\L_M)\e
Combining the first and the third eqs. in (35) and using (17) we have the 
following matching constraint ,                                     
\b\t\a^{-1}_{BL}(\L_M)={5\o 2}\d_1-{3\o 2}\d_2+42.5\e
Approximate unification of gauge couplings at $M_U=10^{18}$ GeV with two sets 
of four fields is found to be possible when the $\D_L-\D_R$ mass difference is 
enhanced but remains within an acceptable limit corresponding to $\rho=
M_{\D_L}/M_{\D_R}=1.6$. The individual masses and values of coupling constants 
at $\mu=M_U=10^{18}$ GeV and $\mu=\L_M=10^{11}$ GeV are found to be,
\[M_{\D_L}=7.8\times 10^{11}{\rm GeV},\ \ 
M_{\D_R}=4.7\times 10^{11}{\rm GeV}\n\]                 
\bm\b M_\xi=5.7\times 10^{11}{\rm GeV},\ \ M_\s=3.3\times 10^{12}{\rm GeV}\e                              
\[\t g_{2L}(M_U)=\t g_{2R}(M_U)=0.640,\ \ \t g_{BL}(M_U)=0.616\n\]                                     
\[\t g_{3C}(M_U)=0.643\pm 0.007,\ \ \t g_6(M_U)=0.636\n\]                                     
\[\t g_{2L}(\L_M)=\t g_{2R}(\L_M)=0.740,\ \ \t g_{BL}(\L_M)=0.508\n\]                                  
\b\t g_{3C}(\L_M)=0.745\pm 0.007\e\em                                   
It is to be noted that the masses of $\D_L$ and $\D_R$ are constrained by              
spontaneous breaking of the left-right discrete symmetry and the 
$SU(2)_R\times U(1)_{B-L}$ gauge symmetry in $G_{2213}$, but there are no such            
constraints on the masses of $\xi$ and $\s$ fields. In no case the mass of any 
of the four fields should be widely different from $\L_M$. From such 
considerations the mass $M_\s=33\L_M$ in (37a) may be near the maximally 
permitted value. However, if there are more than two sets of degenerate 
$\s$-condensates in the model, its mass is likely to decrease. The evolution 
of gauge couplings from $M_Z$ to $M_U$ through $\L_M$ is presented in Fig.2
where threshold effects at lower and intermediate scale are also exhibited. 
\begin{figure}
\begin{center}
\leavevmode
\hbox{\epsfxsize=4in
\epsfysize=4in
\epsffile{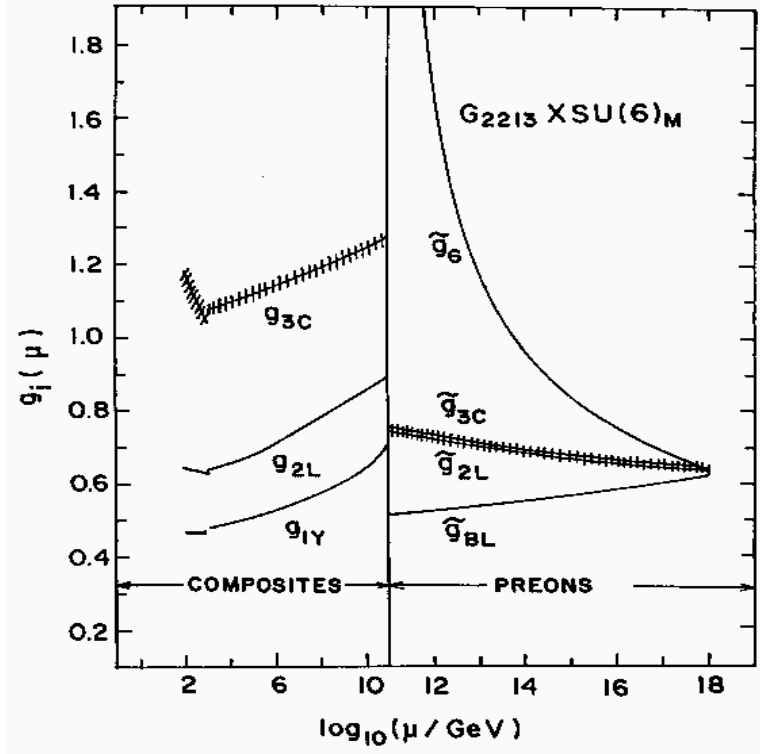}}
\end{center}
\caption{Same as Fig.1 but for the left-right symmetric preonic gauge group 
$SU(2)_L\times SU(2)_R\times U(1)_{B-L}\times SU(3)_C\times SU(6)_M$ with two 
sets of four fields, $\D_L$, $\D_L$, $\xi$ and $\s$, as described in the text 
and for 60\% of mass difference between $\D_L$ and $\D_R$. The $SU(3)_C$ and 
the $SU(2)_L$ couplings follow almost the same trajectory from $\mu=10^{11}$ 
GeV to $10^{18}$ GeV because of one-loop unification in this range in the 
presence of $SU(6)_M$.}
\end{figure}
\subsection{$G_P=SU(2)_L\times U(1)_Y\times SU(3)_C\times SU(6)_M$}
Corresponding to this symmetry $i=1Y,2L,3C$, and 6 in (25) and the one-loop 
coefficients are                                                   
\[b''_{1Y}={7\times 6\o 10}={21\o 5},\ \ b''_{2L}=-3\times 2+{6\o 2}=-3\n\]
\b b''_{3C}=-3\times 3+6=-3,\ \ b''_6=-3\times 6+6=-12\e
It is interesting to note that                                                
\[b''_{2L}=b''_{3C}=-3\n\]                                        
which implies unification of the preonic gauge couplings of $SU(2)_L$ and 
$SU(3)_C$ at one-loop level for all values of $\mu$ from $\L_M$ to $M_U$ as             
explained in Sec V.B i.e.,                                                  
\[\t g_{2L}(\mu)=\t g_{3C}(\mu),\ \ \mu=\L_M\ \ {\rm to}\ \ M_U\n\]                                       
In order to achieve approximate unification of the gauge                
couplings at $M_U\simeq 10^{18}$ GeV, we need $\t\a^{-1}_6(M_U)\simeq 27-30$. 
Neglecting threshold effects at $\L_M$ gives                                                 
\[\t\a^{-1}_{1Y}(\L_M)=\a^{-1}_1(\L_M)=26.7\n\]
\[\t\a^{-1}_{2L}(\L_M)=\a^{-1}_2(\L_M)=16.1\n\]
\b\t\a^{-1}_{3C}(\L_M)=\a^{-1}_3(\L_M)=7.6\pm 0.6\e
Since
\b{21\o 10\pi}\ln{M_U\o\L_M}=10.8,\ \ {3\o 2\pi}\ln{M_U\o\L_M}=7.7\e
eqs.(25) and (38)-(40) have the predictions,                    
\[\t\a^{-1}_{1Y}(M_U)=26.7-10.8=15.9\n\]
\[\t\a^{-1}_{2L}(M_U)=16.1-7.7=23.8\n\]
\b\t\a^{-1}_{3C}(M_U)=7.6+7.7\pm 0.6=15.3\pm 0.6\e
Thus, starting from the CERN-LEP data at $\mu=M_Z$, including SUSY threshold 
effects, but ignoring the intermediate-scale-threshold corrections at 
$\mu=\L_M$, there is no possibility of unification of gauge couplings at the 
preonic level with $G_{fc}=G_{213}$. When attempt is made to unify the gauge 
couplings including intermediate-scale threshold effects,we note from (41) 
that the corrections on each of $\a^{-1}_1(\L_M)$ and $\a^{-1}_3(\L_M)$ must 
be nearly two times as large as that on $\a^{-1}_2(\L_M)$. Including threshold 
effects,the matching conditions at $\L_M$ are                                                                        
\b\a_i^{-1}(\L_M)+\d_i=\t\a^{-1}_i(\L_M),\ \ i=1Y,2L,3C\e
We have observed that a good approximate unification of gauge couplings is 
possible with two sets of four fields if the $\D_L-\D_R$ mass difference is 
enhanced to correspond to the ratio $M_{\D_R}/M_{\D_L}=3.8$ for the following 
values of the individual masses,                       
\[ M_{\D_L}=1.3\times 10^{11}{\rm GeV},\ \ 
M_{\D_R}=5\times 10^{11}{\rm GeV}\n\]                                 
\b M_\xi=3.4\times 10^{11}{\rm GeV},\ \ M_\s=5.7\times 10^{11}{\rm GeV}\e            
The values of the couplings at $M_U$ and $\L_M$ are,                         
\[\t g_{1Y}(M_U)=0.633,\ \ \t g_{2L}(M_U)=0.633,\n\]                                  
\[\t g_{3C}(M_U)=0.655\pm 0.007,\ \ \t g_6(M_U)=0.633,\n\]                                 
\[\t g_{1Y}(\L_M)=0.546,\ \ \t g_{2L}(\L_M)=0.735,\n\]                                 
\b\t g_{3C}(\L_M)=0.758\pm 0.007\e
The evolution of gauge couplings from $M_Z$ to $M_U$ are shown in Fig.3
where the approximate unification at $M_U$ and the one-loop unification of 
$\t g_{2L}(\mu)$ and $\t g_{3C}(\mu)$ for $\mu=\L_M$ to $M_U$ are clearly              
exhibited. Nearly 70\% difference between the $SU(3)_C$-gauge couplings of 
composites and preons compensated by threshold effects at $\L_M$ is found to 
exist in this model. The corresponding differences between the $SU(2)_L$ and 
the $U(1)_Y$ gauge couplings are noted to be nearly 20\% and 27\%, 
respectively.                                  
\begin{figure}
\begin{center}
\leavevmode
\hbox{\epsfxsize=4in
\epsfysize=4in
\epsffile{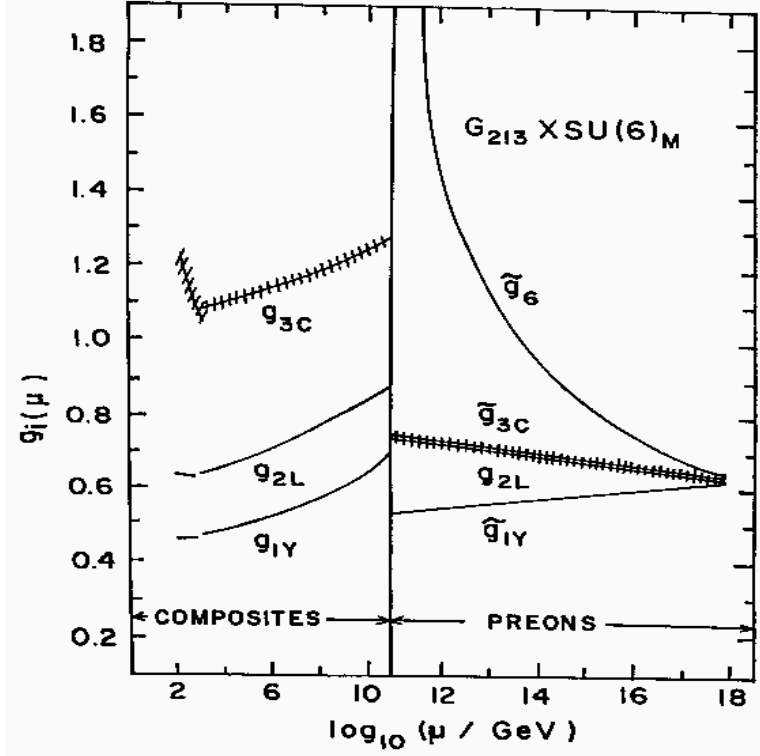}}
\end{center}
\caption{Same as Fig.2 but for the preonic gauge symmetry              
$SU(2)_L\times U(1)_Y\times SU(3)_C\times SU(6)_M$ and 
$M_{\D_R}/M_{\D_L}=3.8$ and two sets of four fields.}
\end{figure}
\subsection{$G_P=SU(2)_L\times U(1)_R\times U(1)_{B-L}\times 
SU(3)_{3C}\times SU(6)_M$}
In this case $i=1R,BL,2L,3C$ and 6 and the one-loop coefficients are                                                                           
\[b''_{1R}={6\o 2}=3,\ \ b''_{BL}=6\n\]
\[b''_{2L}=-3\times 2+{6\o 2}=-3,\ \ b''_{3C}=-3\times 3+6=-3\n\]
\b b''_6=-3\times 6+6=-12\e
As in the cases of $G_{fc}=G_{213}$ and $G_{2213}$, we find $b''_{2L}=
b''_{3C}=-3$ in (45) signifying one-loop unification of preonic gauge 
couplings of $SU(2)_L$ and $SU(3)_C$ over the mass range $\mu=\L_M$ to $M_U$. 
The matching conditions for gauge couplings at $\L_M$ are                                      
\bm\b{1\o\a_1(\L_M)}+\d_1={3\o 5}{1\o\t\a_{1R}(\L_M)}+
{2\o 5}{1\o\t\a_{BL}(\L_M)}\e
\b{1\o\a_2(\L_M)}+\d_2={1\o\t\a_{2L}(\L_M)}\e
\b{1\o\a_3(\L_M)}+\d_3={1\o\t\a_{3C}(\L_M)}\e\em
It is to be noted that one of the gauge couplings in the R.H.S of (46a), 
namely, $\t\a_{1R}(\L_M)$ or $\t\a_{BL}(\L_M)$, appear to remain undetermined.             
But in unified theories, once any of the coupling constant is known at $M_U$, 
the unification constraint gives other gauge couplings at that scale,                                                                  
\[\t\a_{2L}(M_U)=\t\a_{1R}(M_U)=\t\a_{BL}(M_U)=\t\a_{3C}(M_U)\n\]                      
The knowledge of RGEs then determines the values of hitherto unknown couplings 
at lower scales, $\mu<M_U$, With two sets of four fields, we obtain 
$\d_1=16.1$, $\d_2=7.7$ and $\d_3=15.1$ and all the four gauge couplings close 
to one another while satisfying approximate one loop unification, $g_2(\mu)=
g_3(\mu)$, for all $\mu$ from $M_U$ to $\L_M$. The masses of the four fields 
are,                               
\bm\[M_{\D_L}=10^{11}{\rm GeV},\ \ M_{\D_R}=4.4\times 10^{11}{\rm GeV}\n\]                                 
\b M_\xi=5\times 10^{11}{\rm GeV},\ \ M_\s=7.3\times 10^{11}{\rm GeV}\e
The gauge couplings at $M_U$ and $\L_M$ are computed as                              
\[\t g_{1R}(M_U)=\t g_{BL}(M_U)=0.630,\ \ \t g_{2L}(M_U)=0.631\n\]                                
\[\t g_{3C}(M_U)=0.642\pm 0.007,\ \ \t g_6(M_U)=0.641\n\]                                
\[\t g_{1R}(\L_M)=0.563,\ \ \t g_{BL}(\L_M)=0.515\n\]                                
\b\t g_{2L}(\L_M)=0.726,\ \ \t g_{3C}(\L_M)=0.743\pm 0.01\e\em            
Apart from requiring $\rho^{-1}=M_{\D_R}/M_{\D_L}=4.4$, the model also needs             
about 70\% threshold corrections for the $SU(3)_C$ coupling and nearly 20\% 
for the $SU(2)_L$ coupling of composite fields that are introduced by these 
masses.
\subsection{$G_P=SU(2)_L\times SU(2)_R\times SU(4)_{L+R}^C\times SU(6)_M$}
In this case the model possesses left-right discrete symmetry with            
$\t g_{2L}(\mu)=\t g_{2R}(\mu)$ for $\mu=\L_M$ to $M_U$. The one loop 
coefficients are, $b''_{2L}=b''_{2R}=-3$, $b''_{4C}=-6$ and $b''_6=-12$. 
The coupling constants at $\L_M$ are matched using
\[\a^{-1}_2(\L_M)+\d_2=\t\a^{-1}_{2L}(\L_M)=\t\a^{-1}_{2R}(\L_M)\n\]
\[\a^{-1}_3(\L_M)+\d_3=\t\a^{-1}_{4C}(\L_M)\n\]
\[\a^{-1}_1(\L_M)+\d_1={3\o 5}\t\a^{-1}_{2R}(\L_M)+
{2\o 5}\t\a_{4C}^{-1}(\L_M)\]
We have noted that it is impossible to achieve even a roughly approximate 
unification of gauge couplings with the above matching conditions unless 
the number of $\D_L$, $\D_R$, $\xi$ and $\s$ fields are unusually large and 
their masses are widely different from $\L_M$. Thus the flavor-color symmetric 
gauge group $G_{fc}=G_{224}$ is unrealistic.                                                                  
\section{PROSPECTS OF SU(4)-METACOLOR}
In this section, assuming the metacolor gauge symmetry to be $SU(4)_M$, we 
explore possible forms of flavor-color gauge symmetry $G_{fc}$ which could 
unify the relevant gauge couplings at $M_U$, or near the Planck scale. We 
follow  strategies similar to those explained in Sec.V.
\subsection{$G_P=SU(2)_L\times U(1)_R\times U(1)_{B-L}\times 
SU(3)_C\times SU(4)_M$}
With $G_{fc}=G_{2113}$ and $G_M=SU(4)_M$, the one loop coefficients in the 
RGEs of (25) are, $b''_{1R}=2$, $b''_{BL}=4$, $b''_{2L}=-4$, $b''_{3C}=-5$, 
$b''_4=-6$. The matching conditions at $\mu=\L_M$ are given by (46a)-(46c). 
Although one of the gauge couplings, $\t\a_{1R}(\L_M)$ or $\t\a_{BL}(\L_M)$, 
is not determined by the matching conditions, this does not pose a problem in 
studying unification as explained in Sec.V.D. For the sake of simplicity we 
use $\t\a_{1R}(M_U)=\t\a_{BL}(M_U)$ at $M_U=10^{18}$ GeV. Unlike the case of 
$SU(6)_M$, where approximate unification was impossible under a small mass 
difference of 20\% between $M_{\D_L}$ and $M_{\D_R}$, we find that with 
$SU(4)_M$, the gauge group achieves a good approximate unification with gaps 
between the gauge couplings closing in gradually as we approach $\mu=M_{Pl}$. 
The values of masses of the two sets of four fields, needed for approximate 
unification are                                                   
\[M_{\D_L}=5.37\times 10^{10}{\rm GeV},\ \ M_{\D_R}=6.44\times 
10^{10}{\rm GeV}\n\]  
\b M_\s=7.37\times 10^{11}{\rm GeV},\ \ M_\xi=10^{11}{\rm GeV}\e  
where $M_{\D_R}/M_{\D_L}=1.2$. In Table VII we present values of the gauge 
couplings at three different mass scales $\mu=10^{11}$ GeV, $10^{18}$ GeV 
and $10^{19}$ GeV. The evolution of the gauge couplings of the effective gauge 
theories for preons, and quarks and leptons are presented in Fig.4 which
exhibits a clear tandency of the preonic gauge couplings to converge near 
$\mu=M_{Pl}$. The remaining small differences among the couplings at $M_{Pl}$ 
are expected to be filled up by gravitational corrections. One remarkable 
feature of this model is that the difference between the $SU(3)_C$ couplings 
of composite fields and preons is negligible whereas that between the 
$SU(2)_L$ couplings is only 14\%.
\begin{figure}
\begin{center}
\leavevmode
\hbox{\epsfxsize=4in
\epsfysize=4in
\epsffile{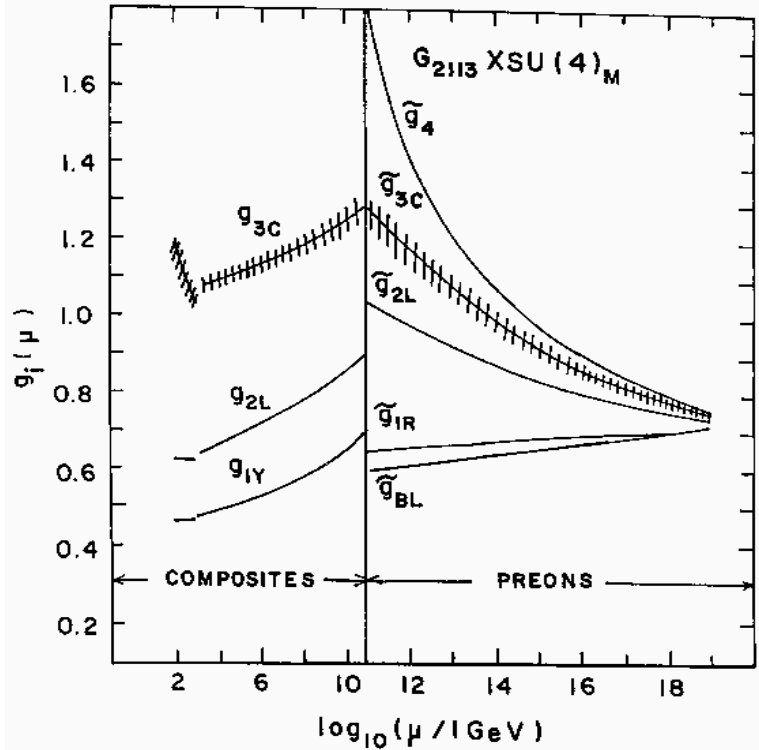}}
\end{center}
\caption{Same as Fig.1 but for the preonic gauge symmetry            
$SU(2)_L\times U(1)_R\times U(1)_{B-L}\times SU(3)_C\times SU(4)_M$ with two 
sets of three fields, $\D_L$, $\D_R$ and $\s$, and $M_{\D_R}/M_{\D_L}=1.2$.}
\end{figure}
\subsection{$G_P=SU(2)_L\times U(1)_Y\times SU(3)_C\times SU(4)_M$}
In the notation of eq.(25), the one loop coefficients are $b''_{1Y}=14/5$,  
$b''_{2L}=-4$, $b''_{3C}=-5$, and $b''_4=-6$. The matching conditions are 
given by eq.(42). Restricting the difference between $M_{\D_L}$ and $M_{\D_R}$ 
to atmost 20\% we find that approximate unification of gauge coupling at 
$\mu=10^{18}-10^{19}$ GeV is impossible. In Table VIII we present values of 
the gauge couplings $\mu=10^{19}$ GeV, $10^{18}$ GeV and $10^{11}$ GeV. Rather 
larger difference between the gauge couplings are found to be contradicting 
the idea of unification. However, we note that the coupling constants can 
unify at $M_U=10^{18}$ GeV only if the $\D_L-\D_R$ mass difference is allowed 
to be larger with $\rho^{-1}=M_{\D_R}/M_{\D_L}\cong 2.9$ corresponding to the 
following values of individual masses,                                     
\[M_{\D_L}=1.7\times 10^{10}{\rm GeV},\ \ 
M_{\D_R}=5.1\times 10^{10}{\rm GeV}\n\]                             
\b M_\xi=1.96\times 10^{11}{\rm GeV},\ \ 
M_\s=7.37\times 10^{11}{\rm GeV}\e
Such a unification of gauge couplings and their evolution down to the $Z$-mass 
are presented in Fig.5. A very attractive feature of this model is that it
needs almost negligible difference between $\t g_{3C}(\L_M)$ and 
$g_{3C}(\L_M)$ and also between $\t g_{1Y}(\L_M)$ and $g_{1Y}(\L_M)$.The model 
is found to require nearly 25\% threshold correction on $SU(2)_L$ coupling of 
composites which is provided by the masses of two sets of four fields given in 
(49).                      
\begin{figure}
\begin{center}
\leavevmode
\hbox{\epsfxsize=4in
\epsfysize=4in
\epsffile{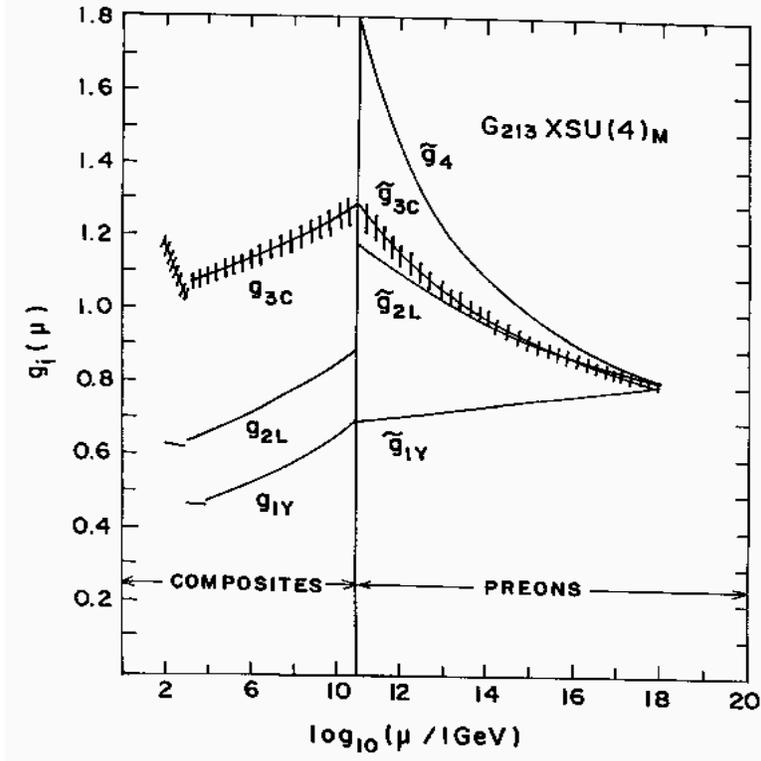}}
\end{center}
\caption{Same as Fig.3 but for the preonic gauge symmetry 
$SU(2)_L\times U(1)_Y\times SU(3)_C\times SU(4)_M$ and 
$M_{\D_R}/M_{\D_L}=2.9$.}                                  
\end{figure}
\subsection{Difficulties with Other Flavor-Color Symmetries}
The flavor-color groups investigated in VI.A-VI.B are the most successful ones 
in the presence of $SU(4)_M$. Difficulties faced with other symmetries are 
summarised as mentioned here. For $G_P=G_{214}\times SU(4)_M$, $b''_{2L}=-4$, 
$b''_{1R}=2$, $b''_{4C}=-8$ and $b''_4=-6$. Although the masses of two sets of 
all four fields needed for unification near $M_{Pl}$ are reasonable, with 
$M_{\D_L}/M_{\D_R}=1.2$, we find,
\[\t g_{4C}(\mu)>\t g_4(\mu),\ \ \mu=10^{11}-10^{14}{\rm GeV}\n\]              
showing that $\t g_4(\mu)$ is no longer the highest coupling near $\mu=\L_M$            
responsible for binding the preons. This is against the basic assumption of 
the model. For $G_P=G_{2213}\times SU(4)_M$, $b''_{BL}=4$, $b''_{2L}=
b''_{2R}=-4$, $b''_{3C}=-5$, and $b''_4=-6$. With two sets of four fields, the 
masses of $\D_L$ and $\D_R$ needed for approximate unification are nearly two 
orders lighter and those of $\xi$ and $\s$ one order heavier than $\L_M$. For 
$G_P=G_{224}\times SU(4)_M$, either the number of some of the four types of 
fields are unusually large, or some of the masses are 5-6 orders different 
from $\L_M$. Because of such undesirable features, $G_{fc}=G_{214}$, 
$G_{2213}$ or $G_{224}$ are unacceptable in the presence of $SU(4)_M$.
\section{SUMMARY AND CONCLUSION}
We have used the CERN-LEP measurements at $M_Z$ to study unity of forces and 
preonic gauge symmetries of the type $G_P=G_{fc}\times G_M$ in the 
scale-unifying preon model [12] which serves to provide a unified origin of 
the diverse mass scales and an explanation of family replication. Threshold 
effects form an important and essential part of gauge-coupling 
renormalization. Neglecting these effects has led to $G_P=G_{214}\times 
SU(5)_M$ as the only successful gauge symmetry of the preonic effective 
Lagrangian[15]. In this analysis, threshold effects are found to play a 
crucial role in determining the unification of forces near the Planck scale 
and, consequently, the gauge symmetry $G_P$ with new possibilities for 
$G_{fc}$ and $G_M=SU(6)_M$ or $SU(4)_{M}$.
\par 
With $SU(6)_M$ as metacolor gauge group, the most attractive possibility of 
flavor-color symmetry is found to be $G_{fc}=G_{214}$ for which a good 
approximate unification of gauge couplings occurs at $\mu=M_{Pl}=
10^{19}$ GeV with only 10\% to 20\% mass difference between $\D_L(3,1,10^C)$ 
and $\D_R(1,3,10^{*C})$ and the model needs just the minimal set of fields, 
$\D_L+\D_R$ and $\bar\D_L+\bar\D_R$, which are essential from considerations 
of left-right symmetry, preservation of SUSY down to the TeV scale, and 
spontaneous symmetry breaking of $G_{214}$ to the standard model gauge group 
at $\L_M$.
\par 
For the next attractive possibility with $SU(6)_M$ corresponding to the 
left-right symmetric gauge group $G_{fc}=G_{2213}$, two sets of all the four 
fields, $\D_L$, $\D_R$, $\xi$ and $\s$, are needed and an approximate 
unification of gauge couplings is possible for acceptable value of the mass 
ratio, $\rho=M_{\D_L}/M_{\D_R}=1.6$, and $M_\xi=5.7\times 10^{11}$ GeV 
provided $M_\s=3.3\times 10^{12}$ GeV. Unification of gauge couplings is also 
observed with the standard model gauge group $G_{fc}=G_{213}$ and similar 
threshold effects with two sets of four fields provided the mass ratio 
$M_{\D_R}/M_{\D_L}=3.8$, and the individual masses of these fields are between 
$1.3\times 10^{11}$ GeV to $5.7\times 10^{11}$ GeV.
\par
With $SU(4)_M$ as the metacolor gauge symmetry, two of the  flavor-color gauge 
symmetries, $G_{213}$ and $G_{2113}$, appear to be quite successful in 
achieving good approximate unification of the relevant gauge couplings at 
$M_U=10^{18}$ GeV, and $10^{19}$ GeV, respectively. For $G_{fc}=G_{2113}$, 
the model needs the $\D_L-\D_R$ mass difference within 20\% and two sets of 
three fields with reasonable values of masses near $\L_M$. With the standard 
model gauge group $G_{fc}=G_{213}$ and $G_M=SU(4)_M$, the $\D_L-\D_R$ mass 
ratio needed is found to be such that $M_{\D_R}/M_{\D_L}=3$ and the other 
masses are $M_\xi=2\times 10^{11}$ GeV and $M_\s=7.4\times 10^{11}$ GeV. In 
this case two sets of all the four fields are needed.                                  
\par
All heavy and superheavy masses used in this paper for threshold effects 
refer to bare masses. They are devoid of wave function renormalisation effects 
which have been shown to be cancelled out by two-loop effects [20]. We assure 
that the bare masses are enough to produce threshold effects needed for 
new gauge symmetries. The cancellation 
observed in ref.[20] does not affect the results and conclusions of this 
analysis.
\par
One of the most challenging problems is to derive the preonic            
model with one of the choices for the metacolor and                   
flavor-color gauge symmetry, mentioned above, from a string theory.             
Also one of the major issues is to address some of the dynamical                
assumptions of the model as regards the  preferred 
directions of symmetry breaking and the saturation of the            
composite spectrum, mentioned in the introduction [14,15]. In the 
absence of a derivation of the model from a deeper theory, apart 
from a number of unproven assumptions, the possible presence of 
more than one flavor-color symmetry groups above $\mu=\L_M$ has an
arbitrariness similar to SUSY $SO(10)$ with different possibilities 
for intermediate gauge symmetries. In spite of present theoretical 
limitations, the preonic approach seems promising because it is 
most economical and explains certain basic issues [12-15], by 
utilising primarily symmetries of the underlying theory and 
general results such as the Witten index theorem, rather than 
detailed dynamics. A crucial test of the model hinges on the 
detection of vectorial quarks and leptons with masses near 1-2 TeV.   
At present, there is no compelling evidence that quarks and leptons 
are composite as proposed in the model, although some 
possible signature has been investigated [16]. 
\acknowledgments
The author acknowledges fruitful collaboration with Professor            
J.C.Pati and Dr.K.S.Babu during the initial stage of this work.
He would also like to thank Professor Pati for very useful         
comments and suggestions and to Dr.Babu for useful discussions. This 
work was partially suported by the grant of the project No.SP/S2/K-09/91 
from the Department of Science and Technology,New Delhi. 
                                    
\mediumtext
\begin{table} 
\caption{Light and heavy spectrum in the scale-unifying preon                  
model and their quantum numbers under $G_{224}$ and $G_{213}$.}
\begin{tabular}{ccc}
Particle type and&Particle type&$G_{213}$-\\
$G_{214}$-quantum nos.&under standard&quantum nos.\\
&model&\\
\tableline
L.H.quarks and leptons&$(u, d)_L$, $(c, s)_L$, $(t, b)_L$&$(2, 1/6, 3)$\\             
$q_L^{e, \mu, \tau}(2_L, 1, 4^*_C)$&$(\nu_e, e)_L$, $(\nu_\mu, \mu)_L$, 
$(\nu_\tau, \tau)_L$&$(2, -1/2, 1)$\\ 
R.H.quarks and leptons&$u_R$, $c_R$, $t_R$&$(1, 2/3, 3)$\\
$q_R^{e, \mu, \tau}(1, 2_R, 4^*_C)$&$d_R$, $s_R$, $b_R$&$(1, -1/3, 3)$\\         
&$e_R$, $\mu_R$, $\tau_R$&$(1, -1, 1)$\\            
&$\nu_{e_R}$, $\nu_{\mu_R}$, $\nu_{\tau_R}$&$(1, 0, 1)$\\ 
L.H.vectorial quarks and&$(U,D)_{L,R}$&$(2, 1/6, 3)$\\
leptons: $Q_{L,R}(2_L, 1, 4^*)$&$(N,E)_{L,R}$&$(2, -1/2, 1)$\\ 
R.H.vectorial quarks and&$(U', D')_{L,R}$&$(1, 2/3, 3)$\\                
leptons $Q'_{L,R}(1, 2_R, 4^*)$&$(N',E')_{L,R}$&$(1, -1/3, 1)$\\ 
Bidoublet of Higgs&$h_u$&$(2, 1/2, 1)$\\              
scalars: $\phi (2, 2, 1)$&$h_d$&$(2, -1/2, 1)$\\ 
Minimal sets of heavy Higgs&&\\                                                   
$\D_L^{1, 2}(3, 1, 10^C)$, $\D_R^{1, 2}(1, 3, 10^{*C})$&see Table V&
see Table V\\ 
Other sets of heavy Higgs&&\\
$\xi^{1, 2}(2, 2, 15)$, $\s^{1, 2}(1, 1, 15)$&see Table IV&see Table IV\\
\end{tabular}
\end{table}
\mediumtext
\begin{table}
\caption{One-loop $\beta$-function coefficients for particles at                  
lower threshold with their quantum numbers and                     
assigned values of masses used for computation of                  
threshold effects.}
\begin{tabular}{cccccc}
Particle type&$G_{213}$-&Mass&$b_1^\a$&$b_2^\a$&$b_3^\a$\\
&quantum nos.&(GeV)&&&\\
\tableline
L.H. top $t_L$&$(2, 1/6, 3)$&175&$1/30$&1&$1/3$\\
R.H. top $t_R$&$(1, 2/3, 3)$&175&$8/15$&0&$1/3$\\
$u$-type Higgs $h_u$&$(2, 1/2, 1)$&120&$1/10$&$1/6$&0\\                
$d$-type Higgs $h_d$&$(2, -1/2, 1)$&250&$1/10$&$1/6$&0\\               
L.H.vectorial quark&&&&&\\
doublets $(U,D)_{L,R}$&$(2, 1/6, 3)$&500&$2/15$&2&$4/3$\\
R.H vectorial $u$-type&&&&&\\                                              
quarks $U'_{L,R}$&$(1, 2/3, 3)$&500&$16/15$&0&$2/3$\\
R.H.vectorial $d$-type&&&&&\\                                                           
quarks $D'_{l,R}$&$(1, -1/3, 3)$&500&$4/15$&0&$2/3$\\
L.H.vectorial lepton&&&&&\\                                                     
doublets $(N,E)_{L,R}$&$(2, -1/2, 1)$&100&$2/5$&$2/3$&0\\               
R.H.vectorial charged&&&&&\\                                                          
leptons $E'_{L,R}$&$(1, -1, 1)$&100&$4/5$&0&0\\            
\end{tabular}
\end{table}
\narrowtext
\begin{table}
\caption{Same as Table II but for superpartners only, and the number of 
squarks and sleptons correspond to summing over three flavors.}
\begin{tabular}{cccccc}
Particle type&$G_{213}$-&Mass&$b_1^\a$&$b_2^\a$&$b_3^\a$\\
&quantum nos.&(GeV)&&&\\                  
\tableline
gluino&(1, 0, 8)&150-200&0&0&23\\         
wino&(3, 0, 1)&100-150&0&$4/3$&0\\             
L.H.slepton doublets&$(2, -1/2, 1)$&500-1500&$3/10$&$1/2$&0\\             
R.H.charged sleptons&$(1, -1, 1)$&500-1500&$3/5$&0&\\                 
L.H.squark doublets&$(2, 1/6, 3)$&500-1500&$1/10$&$3/2$&1\\           
R.H.$u$-type squarks&$(1, 2/3, 3)$&500-1500&$4/15$&0&$1/2$\\        
R.H.$d$-type squarks&$(1, 1/3, 3)$&500-1500&$1/5$&0&$1/2$\\           
$u$-type Higgsino&$(2, 1/2, 1)$&100-300&$1/5$&$1/3$&0\\         
$d$-type Higgsino&$(2, -1/2, 1)$&100-300&$1/5$&$1/3$&0\\          
\end{tabular}
\end{table}
\narrowtext
\begin{table}
\caption{One-loop $\beta$-function coefficients for the components of             
$G_{224}$-Higgs supermultiplet $\xi (2, 2, 15)$ near the metacolor              
scale under the standard gauge group $G_{213}$.}                         
\begin{tabular}{ccccccc}
$G_{213}$&$b_1^\a$&$b_2^\a$&$b_3^\a$&$\sum b_1^\a$&$\sum b_2^\a$&
$\sum b_3^\a$\\
submultiplet&&&&&&\\
\tableline
$\xi_1(2, 1/2, 1)$&$3/10$&$1/2$&0&&&\\                                         
$\xi_2(2, -1/2, 1)$&$3/10$&$1/2$&0&&&\\                                         
$\xi_3(2, -1/6, 3)$&$1/10$&$3/2$&1&&&\\                                         
$\xi_4(2, -7/6, 3)$&$49/10$&$3/2$&1&&&\\                                         
$\xi_5(2, -7/6, \overline{3})$&$49/10$&$3/2$&1&$77/5$&15&16\\             
$\xi_6(2, 1/6, \overline{3})$&$1/10$&$3/2$&1\\                                         
$\xi_7(2, 1/2, 8)$&$24/10$&4&6&&&\\                                         
$\xi_8(2, -1/2, 8)$&$24/10$&4&6&&&\\                                         
\end{tabular}
\end{table}
\narrowtext
\begin{table}
\caption{Same as Table IV but for $G_{224}$-multiplets,                      
$\D_L(3, 1, 10^C)$, $\D_R (1, 3, 10^{*C})$, and $\s (1, 1, 15)$.}                             
\begin{tabular}{ccccccc}
$G_{213}$&$b_1^\a$&$b_2^\a$&$b_3^\a$&$\sum b_1^\a$&$\sum b_2^\a$&
$\sum b_3^\a$\\
submultiplet&&&&&&\\
\tableline
$\D_{R_1} (1, 1, 1)$&$3/5$&0&0&&&\\                                         
$\D_{R_2} (1, 2, 1)$&$1/5$&0&0&&&\\                                       
$\D_{R_3} (1, 1/3, \overline{3})$&$1/5$&0&$1/2$&&&\\                                         
$\D_{R_4} (1, 2/3, \overline{3})$&$4/5$&0&$1/2$&&&\\                                        
$\D_{R_5} (1, -4/3, \overline{3})$&$16/5$&0&$1/2$&$78/5$&0&9\\             
$\D_{R_6} (1, 1/3, \overline{6})$&$2/5$&0&$5/2$&&&\\                                       
$\D_{R_7} (1, -2/3, \overline{6})$&$8/5$&0&$5/2$&&&\\                                       
$\D_{R_8} (1, 4/3, \overline{6})$&$32/5$&0&$5/2$&&&\\                                       
$\D_{L_1} (3, 1, 1)$&$9/5$&2&0&&&\\                                         
$\D_{L_2} (3, 1/3, \overline{3})$&$3/5$&6&$3/2$&$18/5$&20&9\\             
$\D_{L_3} (3, -1/3, 6)$&$6/5$&12&$15/2$&&&\\                                      
$\s_1(1, -2/3, 3)$&$4/5$&0&$1/2$&&&\\                                      
$\s_2(1, 2/3, 3)$&$4/5$&0&$1/2$&$8/5$&0&4\\              
$\s_3(1, 0, 8)$&0&0&3&&&\\
\end{tabular}
\end{table}
\narrowtext
\begin{table}
\caption{Gauge couplings at different mass scales in the presece of two sets 
of relevant Higgs superfields for the preonic symmetry $G_P=SU(2)_L\times 
U(1)_R\times SU(4)_{L+R}^C\times SU(6)_M$, with $M_{\D_L}=3.5\times 10^{11}$ 
GeV and $\rho=M_{\D_L}/M_{\D_R}=1.1$. Note that the four gauge couplings 
converge within 4\% as $\mu$ approaches $10^{19}$ GeV.}                
\begin{tabular}{ccccc}
Mass scale ($\mu$)&$\t g_{1R}(\mu)$&$\t g_{2L}(\mu)$&$\t g_{4C}(\mu)$&
$\t g_6(\mu)$\\ (GeV)&&&&\\ \tableline
$10^{19}$&0.580&0.617&$0.624\pm 0.007$&0.613\\              
$10^{18}$&0.570&0.628&$0.646\pm 0.007$&0.658\\              
$10^{11}$&0.527&0.722&$0.927\pm 0.007$&---\\                
\end{tabular}
\end{table}
\narrowtext
\begin{table} 
\caption{Values of gauge couplings at different mass scales obtained using two 
sets of relevant Higgs superfields for the preonic gauge symmetry $G_P=
SU(2)_L\times U(1)_{I_{3R}}\times U(1)_{B-L}\times SU(3)_C\times SU(4)_M$  
with $M_{\D_L}=5.3\times 10^{10}$ GeV, $M_{\D_R}=6.4\times 10^{10}$ GeV,            
$M_\s=7.3\times 10^{11}$ GeV and $\rho^{-1}=M_{\D_R}/M_{\D_L}=1.2$.}
\begin{tabular}{cccccc}
Mass scale ($\mu$)&$\t g_{2L}(\mu)$&$\t g_{3C}(\mu)$&
$\t g_{1R}(\mu)$&$\t g_{BL}(\mu)$&$\t g_4(\mu)$\\ 
(GeV)&&&&&\\ \tableline
$10^{19}$&0.731&$0.753\pm 0.013$&0.720&0.732&0.764\\             
$10^{18}$&0.755&$0.792\pm 0.013$&0.710&0.710&0.806\\              
$10^{11}$&1.034&$1.32\pm 0.05$&0.647&0.598&1.80\\
\end{tabular}
\end{table}
\mediumtext
\begin{table}
\caption{Gauge couplings at different mass scales but for $G_P=SU(2)_L\times 
U(1)_Y\times SU(3)_C\times SU(4)_M$. Here $M_\s=7.3\times 10^{11}$ GeV 
throughout. For the case(a) $M_{\D_L}=1.3\times 10^{10}$ GeV, $M_{\D_R}=
1.6\times 10^{10}$ GeV, $M_{\D_R}/M_{\D_L}=1.2$, and $M_\xi=4\times 10^{11}$ 
GeV, but $M_{\D_L}=1.7\times 10^{10}$ GeV, $M_{\D_R}=5.1\times 10^{10}$ GeV, 
$M_{\D_R}/M_{\D_L}=2.9$, and $M_\xi=1.96\times 10^{11}$ GeV for the case (b).}
\begin{tabular}{cccccc}
&Mass scale ($\mu$)&$\t g_1(\mu)$&$\t g_{2L}(\mu)$&$\t g_{3C}(\mu)$&
$\t g_4(\mu)$\\ &(GeV)&&&&\\ \tableline
(a)&$10^{18}$&$0.872\pm 0.005$&$0.798\pm 0.008$&$0.808\pm 0.06$&0.859\\              
&$10^{11}$&$0.707\pm 0.004$&$1.149\pm 0.01$&$1.4\pm 0.07$&2.5\\                 
\tableline
(b)&$10^{18}$&$0.802\pm 0.004$&$0.806\pm 0.007$&$0.784\pm 0.015$&0.806\\             
&$10^{11}$&$0.686\pm 0.002$&$1.175\pm 0.01$&$1.287\pm 0.015$&1.80\\               
\end{tabular}
\end{table}
\end{document}